\documentclass[aps,twocolumn,floatfix,superscriptaddress,showpacs,nofootinbib]{revtex4}
\usepackage{graphicx}
\usepackage{amsmath}
\usepackage{amssymb}
\usepackage{bm}

\begin{document}


\title{Numerical Modeling of Coexistence, Competition and Collapse\\
of Rotating Spiral Waves in Three-Level Excitable Media\\ with
Discrete Active Centers and Absorbing Boundaries}
\author{S.~D.~Makovetskiy}%
\email{sdmakovetskiy@mail.ru} \affiliation{Kharkiv National
University of Radio Electronics,\\ 14, Lenin avenue, Kharkiv,
61166, Ukraine}

\date{February 14, 2006}

\begin{abstract}

Spatio-temporal dynamics of excitable media with discrete
three-level active centers (ACs) and absorbing boundaries is
studied numerically by means of a deterministic three-level model
(see S.~D.~Makovetskiy and D.~N.~Makovetskii,
http://arxiv.org/abs/cond-mat/0410460 ), which is a generalization
of the Zykov-Mikhailov model (see Sov. Phys. -- Doklady, 1986,
Vol.31, No.1, P.51) for the case of two-channel diffusion of
excitations. In particular, we revealed some qualitatively new
features of coexistence, competition and collapse of rotating
spiral waves (RSWs) in three-level excitable media under conditions
of strong influence of the second channel of diffusion. Part of
these features are caused by unusual mechanism of RSWs evolution
when RSW's cores get into the surface layer of an active medium
(i.~e. the layer of ACs resided at the absorbing boundary). Instead
of well known scenario of RSW collapse, which takes place after
collision of RSW's core with absorbing boundary, we observed
complicated transformations of the core leading to nonlinear
``reflection'' of the RSW from the boundary or even to birth of
several new RSWs in the surface layer. To our knowledge, such
nonlinear ``reflections'' of RSWs and resulting die hard vorticity
in excitable media with absorbing boundaries were unknown earlier.

\vspace{6pt}

\noindent
\begin{footnotesize}ACM classes: F.1.1~Models of
Computation (Automata), I.6~Simulation and Modeling
(I.6.3~Applications), J.2~Physical Sciences and Engineering
(Chemistry, Physics)
\end{footnotesize}

\end{abstract}

\pacs{05.65.+b, 07.05.Tp, 82.20.Wt}

\maketitle

\section{INTRODUCTION}\label{se:i}

Rotating spiral waves (RSWs) are typical robust self-organized
structures (autowaves) in various dissipative systems, ranging from
chemistry, biology and ecology
\cite{Field-1985,Kapral-1995,Medvinskii-2002} to nonlinear optics
and lasers \cite{Weiss-1999}.

Evolution of these spatio-temporal structures is now a subject of
extensive theoretical studies, experimental investigations and
computer modeling (see, e.g.,
\cite{Lo-2003,Oya-2005,Provata-2003,Szabo-2002,Vanag-2004} and
references therein).

Many mathematical models in this field are based on concept of
excitability \cite{Murray-2002}. An important approach in modern
computer investigations of excitable systems is based on using of
discrete parallel models with local interactions, mostly by
cellular automata (CA).

In the framework of the CA approach
\cite{Bandman-2006,Wolfram-2002}, the three-level representation of
the excitable media states is an effective tool for carrying
computer experiments with multi-particle systems possessing
long-time nonstationary dynamics (see
\cite{SDM-DNM-cond-mat-0410460} and references therein).
Parallelization of updating of the active medium states and
locality of interactions between elementary parts of the medium are
the fundamental principles of the CA approach. Using of this
approach admits to fulfill modeling of large systems by emulation
of parallel matrix transformations.

The essence of three-level representation of excitable medium
\cite{Wiener-1946,Zykov-1986,SDM-DNM-cond-mat-0410460} is as
follows. Each elementary part or active center (AC) of an excitable
system has the single stable {\textit{ground}} state, say
$L_{\mathrm{I}}$, and at least two metastable (upper) states, say
$L_{\mathrm{II}}$ and $L_{\mathrm{III}}$.

The higher metastable state $L_{\mathrm{III}}$ is the
{\textit{excited}} one, it may be reached only after certain hard
perturbation (from neighboring ACs or from external source) of the
ground state $L_{\mathrm{I}}$ of the considered AC. A group of
excited ACs may excite another ACs etc. Due to metastabilty, the
lifetime $\tau_e$ of the excited state $L_{\mathrm{III}}$ is a
finite value, after which every excited AC spontaneously reaches
the intermediate {\textit{refractory}} state $L_{\mathrm{II}}$.

Refractority means ``sleeping'' state at which no excitation of the
considered AC is possible both from neighboring ACs or from an
external source. Moreover, the AC in refractory state can't
participate in excitation of another ACs (lying at
$L_{\mathrm{I}}$). So, after a refractority lifetime $\tau_r$, the
considered AC reaches the ground state $L_{\mathrm{I}}$. Usually
only the cyclic transitions ($L_{\mathrm{I}} \rightarrow
L_{\mathrm{III}} \rightarrow L_{\mathrm{II}} \rightarrow
L_{\mathrm{I}} \rightarrow ...$) are permitted.

The concept of excitability was developed in biology, but it is
widely used now in chemistry and physics. In particular, this
concept have been applied to solve many problems arising in modern
nonlinear optics, laser physics etc.
\cite{Weiss-1993,Staliunas-1995,Weiss-2004,Hagberg-Diss-1994,Giudici-Tredicce-1997,Piwonski-etc,Gomila-2004}

In \cite{SDM-DNM-cond-mat-0410460} we have considered even more
close laser-like analog of excitable system, namely the three-level
active medium of the microwave phonon laser (phaser) with
dipole-dipole interactions between AC. The only significant
difference of phaser medium from the usual excitable one is the
second channel of diffusion of excitations. This system was
experimentally studied in Institute of Radio-Physics and
Electronics (Kharkov, Ukraine)
\cite{DNM-Diss-1983,TPL-2001,arx-selforg-2003,TP-2004,arx04a-destab-2004,UFZh-2006},
demonstrating self-organization, bottlenecked cooperative transient
processes and other nonlinear phenomena under extremely low level
of intrinsic quantum noise (phaser has 15 orders lower intensity of
spontaneous emission in comparison to usual optical-range lasers).

In preceding publications
\cite{SDM-DNM-cond-mat-0410460,SDM-Forum-2005,SDM-BScThesis-2005,SDM-Forum-2006}
we carried out a series of computer experiments on RSWs dynamics in
autonomous excitable systems with various control parameters and
random initial excitations. The most interesting phenomena of
self-organized vorticity observed in computer experiments
\cite{SDM-DNM-cond-mat-0410460,SDM-Forum-2005,SDM-BScThesis-2005,SDM-Forum-2006}
were as follows: (a)~spatio-temporal transient chaos in form of
highly bottlenecked collective evolution of excitations by RSWs
with variable topological charges; (b)~competition of left-handed
and right-handed RSWs with unexpected features, including
self-induced alteration of integral effective topological charge;
(c)~transient chimera states, i.~e. coexistence of regular and
chaotic domains in excitable media; (d)~branching of an excitable
medium states with different symmetry which may lead to full
restoring of symmetry of imperfect starting pattern. Phenomena (a)
and (c) are directly related to microwave phonon laser dynamics
features observed earlier in real experiments at liquid helium
temperatures on corundum crystals doped by iron-group ions
\cite{DNM-Diss-1983,TPL-2001,arx-selforg-2003,TP-2004,arx04a-destab-2004,UFZh-2006}.

In the present paper, we report some qualitatively new features of
coexistence, competition and collapse of RSWs in three-level model
\cite{SDM-DNM-cond-mat-0410460,SDM-Forum-2005,SDM-BScThesis-2005,SDM-Forum-2006}
of excitable media under conditions of strong influence of the
second channel of diffusion of excitations. Part of these features
are caused by unusual mechanism of RSWs evolution when RSW's cores
get into the surface layer of an active medium (i.~e. the layer of
ACs resided at the absorbing boundary). Instead of well known
scenario of RSW collapse, which takes place after collision of
RSW's core with absorbing boundary, we observed complicated
transformations of the core leading to ``reflection'' of the RSW
from the boundary or even to birth of several new RSWs in the
surface layer. To our knowledge, such nonlinear ``reflections'' of
RSWs and resulting die hard vorticity in excitable media with
absorbing boundaries were unknown earlier.

\section{SOME REMARKS ON ZYKOV-MIKHAILOV MODEL AND
ITS TWO-CHANNEL MODIFICATION}\label{se:ii-new05}

The Zykov-Mikhailov (ZM) model \cite{Zykov-1986} is a discrete
parallel mapping with local interactions, which may be defined as
three-level CA \cite{Loskutov-1990}. Let an active discretized
medium ${\mathfrak{M}}_{\mathrm{e}}$ has the form of rectangular
${\mathrm{2D}}$\nobreakdash-\hspace{0pt}lattice. Each cell of the
lattice contains the single three-level AC with coordinates
$(i,j)$, where $(\min(i) = 1)$, $(\min(j) = 1)$. All the ACs in the
${\mathfrak{M}}_{\mathrm{e}}$ are identical, and they interact by
the same set of rules (ZM model, as well as most of the CA models,
is homogeneous and isotropic in the von~Neumann sense
\cite{Neumann-1966}). Each AC has the single stable
{\textit{ground}} state $L_{\mathrm{I}}$, and two metastable states
$L_{\mathrm{II}}$ and $L_{\mathrm{III}}$.

The upgrade of states $S_{ij}^{(n)} \equiv S^{(n)}(i,j)$ of all ACs
(i.~e. their intralevel or interlevel transitions) are carrying out
synchronously at each step $0 < n \leq N$ during the system
evolution. All intralevel ($L_{\mathrm{I}} \rightarrow
L_{\mathrm{I}}$, $L_{\mathrm{II}} \rightarrow L_{\mathrm{II}}$,
$L_{\mathrm{III}} \rightarrow L_{\mathrm{III}}$) and some of
interlevel (namely $L_{\mathrm{I}} \rightarrow L_{\mathrm{III}}$,
$L_{\mathrm{III}} \rightarrow L_{\mathrm{II}}$, $L_{\mathrm{II}}
\rightarrow L_{\mathrm{I}}$) transitions are permitted. The final
step $N$ may be either predefined or it will be searched during CA
evolution as time (i.e. quantity of discrete steps) for reaching an
attractor of this CA. The conditions of upgrade
\cite{Zykov-1986,Loskutov-1990} depend both on $S_{ij}^{(n-1)}$ and
on $S_{i'j'}^{(n-1)}$, where $\{i',j'\}$ belongs to certain active
neigborhood of AC at site $(i,j)$.  In the framework of the ZM
model (and in most of other CA models of excitable chemical
systems), diffusion of excitations is possible only by
$L_{\mathrm{I}} \rightarrow L_{\mathrm{III}}$ transition. This is a
single-channel (1C) mechanism of diffusion.

A modification of ZM model was proposed by us
\cite{SDM-DNM-cond-mat-0410460,SDM-Forum-2005,SDM-BScThesis-2005,SDM-Forum-2006}
to adapt it for emulation of some aspects of dynamics of class-B
optical lasers \cite{Weiss-1993,Staliunas-1995,Weiss-2004} and
microwave phonon lasers
\cite{DNM-Diss-1983,TPL-2001,arx-selforg-2003,TP-2004,arx04a-destab-2004,UFZh-2006}.
The relaxational properties of ACs in microwave phonon lasers are
of the same type, as in class-B optical lasers
\cite{Tredicce-1985}, and there are experimental confirmations
\cite{TPL-2001,arx-selforg-2003,TP-2004,arx04a-destab-2004} of
common properties of microwave phonon lasers and class-B optical
lasers.

In order to formulate boundary conditions for a
${\mathfrak{M}}_{\mathrm{e}}$ of finite size (i.~e. when $\max(i) =
M_X, \max(j) = M_Y$), a set ${\mathfrak{M}}_{\mathrm{v}}$ of
virtual cells with coordinates $(i = 0) \vee (j = 0) \vee (i = M_X
+ 1) \vee (j = M_Y + 1)$ may be introduced
\cite{SDM-DNM-cond-mat-0410460,SDM-Forum-2005,SDM-BScThesis-2005,SDM-Forum-2006}.
Each virtual cell contains the unexcitable center having the single
level $L_0$. The surface layer of a bounded active system is
represented by the set ${\mathfrak{M}}_{\mathrm{surf}}$ of ACs with
coordinates $(i = 1) \vee (j = 1) \vee (i = M_X) \vee (j = M_Y)$.
Note that our surface layer is of 1D type because active medium
itself is of 2D type (we stress this circumstance to avoid
misunderstandings).

The original ZM model has, as it was pointed out earlier, the
single channel of diffusion \cite{Zykov-1986,Loskutov-1990}. Such
1C-diffusion models are adequate for chemical reaction-diffusion
systems \cite{Tyson-1985}. In microwave phonon laser active system,
the multichannel diffusion of spin excitations is the typical case,
because it proceeds via (near)-resonant dipole-dipole ($d-d$)
magnetic interactions between paramagnetic ions
\cite{DNM-Diss-1983}. For a three-level system, which is the
simplest microwave phonon laser system, there are 3 possible
channels of resonant diffusion. In the case when $d-d$ interactions
are forbidden at one of three resonant frequencies of a three-level
system, the asymmetric two-channel (2C) diffusion is realized under
conditions of perfect refractority of the intermediate level ---
see
\cite{SDM-DNM-cond-mat-0410460,SDM-Forum-2005,SDM-BScThesis-2005,SDM-Forum-2006}.
Note that asymmetric diffusion is well known in biology (see, e.g.,
the book of D.~A.~Frank-Kamenetzky \cite{Frank-1984}). Recently a
kind of asymmetric diffusion was proposed by N.~Packard and R.~Shaw
\cite{Packard-2004} for a mechanical system.

\section{THREE-LEVEL MODEL OF EXCITABLE SYSTEM WITH TWO-CHANNEL DIFFUSION
(A~MODIFIED ZYKOV-MIKHAILOV MODEL)}\label{se:iii}

\subsection{States of active centers and branches of
evolution operator}\label{subse:iii-A}

Let ${\mathfrak{M}}_{\mathrm{e}}$ is a rectangular 2D lattice
containing $M_X \times M_Y$ three-level ACs. The upgrade of state
$S_{ij}^{(n)}$ of each AC is carried out synchronously during the
system evolution. The excited (upper) level $L_{\mathrm{III}}$ has
the time of relaxation $\tau_e \geq 1$, and the refractory
(intermediate) level $L_{\mathrm{II}}$ has the time of relaxation
$\tau_r \geq 1$. Both $\tau_e$ and $\tau_r$ are integer numbers.
This model of excitable system is a kind of three-level CA. The
original ZM model \cite{Zykov-1986,Loskutov-1990} was formulated as
three-level CA with 1C diffusion.

In our three-level model of excitable system (TLM), the second
channel of diffusion of excitations is included. A detailed
argumentation for using of the 2C diffusion mechanism one can find
in \cite{SDM-DNM-cond-mat-0410460}.

In the TLM, the first channel of diffusion accelerates the
transitions $L_{\mathrm{I}} \rightarrow L_{\mathrm{III}}$ for a
given AC, and the second channel of diffusion (which is absent in
the ZM model) accelerates the transitions $L_{\mathrm{III}}
\rightarrow L_{\mathrm{II}}$. The complete description of AC's
states $S_{ij}^{(n)}$ in such the TLM model includes one type of
global attributes (the phase counters $\varphi_{ij}^{(n)}$) and two
types of partial attributes $u_{ij}^{(n)}$ and $z_{ij}^{(n)}$ for
each individual AC in ${\mathfrak{M}}_{\mathrm{e}}$. Full
description of all AC's possible states is as follows:
\begin{equation}
  S_{ij}^{(n)} \bigl( L_{\mathrm{I}} \bigr) =
  \bigl( \varphi_{ij}^{(n)}, u_{ij}^{(n)} \bigr);
  \label{eq:01}
\end{equation}
\begin{equation}
  S_{ij}^{(n)} \bigl( L_{\mathrm{III}} \bigr) =
  \begin{cases}
    \bigl( \varphi_{ij}^{(n)} \bigr),               & \text{if $n = 0$},\\
    \bigl( \varphi_{ij}^{(n)}, z_{ij}^{(n)} \bigr), & \text{if $n \neq 0$};
  \end{cases}
  \label{eq:02}
\end{equation}
\begin{equation}
  S_{ij}^{(n)} \bigl( L_{\mathrm{II}} \bigr) =
  \bigl( \varphi_{ij}^{(n)} \bigr).
  \label{eq:03}
\end{equation}

In the framework of the TLM model, the phase counters lie in the
interval $\varphi_{ij}^{(n)} \in [0, \tau_e + \tau_r]$. The
following correspondences between $\varphi_{ij}^{(n)}$ and $L_K$
take place (by definition) for all the ACs in
${\mathfrak{M}}_{\mathrm{e}}$ at all steps $n$ of evolution:
\begin{gather}
  \bigl( L_K = L_{\mathrm{I}} \bigr) \Leftrightarrow
  \bigl( \varphi_{ij}^{(n)} = 0 \bigr); \label{eq:04}\\
  \bigl( L_K = L_{\mathrm{III}} \bigr) \Leftrightarrow
  \bigl( 0 < \varphi_{ij}^{(n)} \leq \tau_e \bigr); \label{eq:05}\\
  \bigl( L_K = L_{\mathrm{II}} \bigr) \Leftrightarrow
  \bigl( \tau_e < \varphi_{ij}^{(n)} \leq \tau_e + \tau_r \bigr).
  \label{eq:06}
\end{gather}

These correspondences are of key significance for all
multistep-relaxation models (Reshodko model
\cite{Bogach-1973,Bogach-1979,Reshodko-etc}, ZM model
\cite{Zykov-1986,Loskutov-1990} etc.): there are only three
discrete levels, and relaxation of each AC is considered as
intralevel transitions (or transitions between virtual sublevels).

\begin{widetext}

The evolution of each individual AC proceeds by sequential cyclic
transitions $L_K \rightarrow L_{K'}$ (where $K$ and $K' \in \{
{\mathrm{I}}, {\mathrm{III}}, {\mathrm{II}} \}$), induced by the
Kolmogorov evolution operator $\widehat\Omega$
\cite{Kolmogorov-etc}. In the TLM model, the evolution operator
$\widehat\Omega$ has three orthogonal branches
$\widehat\Omega_{\mathrm{I}}$, $\widehat\Omega_{\mathrm{III}}$ and
$\widehat\Omega_{\mathrm{II}}$, which we call ground, excited and
refractory branches respectively. The choice of the branch at
iteration $n+1$ is dictated only by the global attribute of the AC
at step $n$, namely:

\begin{equation}
  \varphi_{ij}^{(n+1)} = \widehat\Omega\left(\varphi_{ij}^{(n)}\right) =
  \begin{cases}
    \widehat\Omega_{\mathrm{I}}\left(\varphi_{ij}^{(n)}\right),  & \text{if $\varphi_{ij}^{(n)} = 0$},\\
    \widehat\Omega_{\mathrm{III}}\left(\varphi_{ij}^{(n)}\right),& \text{if $0 < \varphi_{ij}^{(n)} \leq \tau_e$},\\
    \widehat\Omega_{\mathrm{II}}\left(\varphi_{ij}^{(n)}\right), & \text{if $\tau_e < \varphi_{ij}^{(n)} \leq \tau_e + \tau_r$}.
  \end{cases}
  \label{eq:07}
\end{equation}

\subsection{Ground branch of the evolution operator}\label{subse:iii-B}

At step $n+1$, the branch $\widehat\Omega_{\mathrm{I}}$ by
definition fulfills operations only over those ACs, which have $L_K
= L_{\mathrm{I}}$ at step $n$. These operations are precisely the
same, as in ZM model \cite{Zykov-1986,Loskutov-1990}, namely:

\begin{equation}
  \varphi_{ij}^{(n+1)} = \widehat\Omega_{\mathrm{I}}\left(\varphi_{ij}^{(n)}\right) =
  \begin{cases}
    0, & \text{if $\bigl( \varphi_{ij}^{(n)} = 0 \bigr) \wedge \bigl( u_{ij}^{(n+1)} < h \bigr)$};\\
    1, & \text{if $\bigl( \varphi_{ij}^{(n)} = 0 \bigr) \wedge \bigl( u_{ij}^{(n+1)} \geq h \bigr)$},
  \end{cases}
  \label{eq:08}
\end{equation}

\vspace{12pt}

\begin{equation}
  u_{ij}^{(n+1)} =
  A_{ij}^{(n)} \left( S_{ij}^{(n)} \right) + D_{ij}^{(n)} \left( S_{i+p,j+q}^{(n)} \right) =
  g u_{ij}^{(n)} + \sum_{p,q} C(p,q)J_{i+p,j+q}^{(n)},
  \label{eq:09}
\end{equation}
where $A_{ij}^{(n)}$ is the accumulating term for the
$u$\nobreakdash-\hspace{0pt}agent, which is an analog of chemical
activator \cite{Zykov-1986,Loskutov-1990}; $D_{ij}^{(n)}$ is the
first-channel diffusion term; $h$ is the threshold for the
$u$\nobreakdash-\hspace{0pt}agent ($h
> 0$); $g$ is the accumulation factor for the
$u$\nobreakdash-\hspace{0pt}agent ($g \in [0,1])$; $C(p,q)$ is the
active neighborhood of the AC at site $(i,j)$; and the
$u$\nobreakdash-\hspace{0pt}agent arrives to ACs with $L_K =
L_{\mathrm{I}}$ only from ACs with $L_K = L_{\mathrm{III}}$ in
$C(p,q)$:
\begin{equation}
  J_{i+p,j+q}^{(n)} =
  \begin{cases}
    1, & \text{if $\bigl( 0 < \varphi_{i+p,j+q}^{(n)} \leq \tau_e \bigr)$};\\
    0, & \text{if $\bigl( \varphi_{i+p,j+q}^{(n)} > \tau_e \bigr) \vee \bigl( \varphi_{i+p,j+q}^{(n)} = 0 \bigr)$}.
  \end{cases}
  \label{eq:10}
\end{equation}

The definition of the diffusion term $D_{ij}^{(n)}$ in Eqn.
(\ref{eq:09}) is very flexible. Apart of well-known neighborhoods
of the Moore ($C(p,q) = C_M(p,q)$) and the von Neumann ($C(p,q) =
C_N(p,q)$) types:

\begin{equation}
  C_M(p,q) =
  \begin{cases}
    1, & \text{if $\Bigl[ \bigl( \, |p| \leq 1 \bigr) \wedge \bigl( \, |q| \leq 1 \bigr) \Bigr] \wedge \bigl( \delta_{p0}\delta_{q0} \neq 1 \bigr)$};\\
    0, & \text{otherwise},
  \end{cases}
  \label{eq:11}
\end{equation}
\begin{equation}
  C_N(p,q) =
  \begin{cases}
    1, & \text{if $\Bigl[ \bigl( \, |r| = 1 \bigr) \oplus \bigl( \, |s| = 1 \bigr) \Bigr]$};\\
    0, & \text{otherwise},
  \end{cases}
  \label{eq:12}
\end{equation}
one can easy define another neighborhoods, e.~g. of the box type
$C_B(p,q)$ or of the diamond type $C_D(p,q)$ \cite{Fisch-1991}
(which are the straightforward generalization of the  Moore
$C_M(p,q)$ and the von Neumann $C_N(p,q)$ neighborhoods), etc.

On the other hand, the ZM definition
\cite{Zykov-1986,Loskutov-1990} of the weight factors $
J_{i+p,j+q}^{(n)}$ (\ref{eq:10}) may be extended out of the binary
set $\{0,1\}$ to model various distance-dependent phenomena within,
e.~g., $C_D(p,q), C_B(p,q)$.

In this work, however, we restricted ourselves by the Moore
neighborhood and by weight factors of the form (\ref{eq:10}).
Another types of neighborhoods, weight factors and some other
modifications of the model will be studied in subsequent papers.

\subsection{Excited branch of the evolution operator}\label{subse:iii-C}

At the same step $n+1$, the branch $\widehat\Omega_{\mathrm{III}}$
fulfills operations only over those ACs, which have $L_K =
L_{\mathrm{III}}$ at step $n$:
\begin{equation}
  \varphi_{ij}^{(n+1)} =
  \widehat\Omega_{\mathrm{III}}\left( \varphi_{ij}^{(n)} \right)=
  \begin{cases}
    \varphi_{ij}^{(n)} + 1, & \text{if $\left[ \bigl( 0 < \varphi_{ij}^{(n)} < \tau_e \bigr) \wedge \bigl( z_{i,j}^{(n+1)} < f \bigr) \right] \vee \bigl( \varphi_{ij}^{(n)} = \tau_e \bigr)$};\\
    \varphi_{ij}^{(n)} + 2, & \text{if $\bigl( 0 < \varphi_{ij}^{(n)} < \tau_e \bigr) \wedge \bigl( z_{i,j}^{(n+1)} \geq f \bigr)$}.
  \end{cases}
  \label{eq:13}
\end{equation}
\begin{equation}
  z_{ij}^{(n+1)} =
  \overline{D}_{ij}^{\:(n)} \left( S_{i+p,j+q}^{(n)} \right) =
  \sum_{p,q} C(p,q)Q_{i+p,j+q}^{(n)}.
  \label{eq:14}
\end{equation}
where $\overline{D}_{ij}^{\:(n)}$ is the second-channel diffusion
term; $f$ is the threshold for the
$z$\nobreakdash-\hspace{0pt}agent ($f>0$), and we assume that
$z$\nobreakdash-\hspace{0pt}agent arrives to excited AC ($L_K =
L_{\mathrm{III}}$) only from those ACs, which have $L_K =
L_{\mathrm{I}}$ in $C(p,q)$:
\begin{equation}
  Q_{i+p,j+q}^{(n)} =
  \begin{cases}
    1, & \text{if $\varphi_{i+p,j+q}^{(n)} = 0$};\\
    0, & \text{if $\varphi_{i+p,j+q}^{(n)} \neq 0$}.
  \end{cases}
  \label{eq:15}
\end{equation}

One can see from (\ref{eq:14}) that the
$z$\nobreakdash-\hspace{0pt}agent does not accumulate during
successive iterations. In other words, the branch
$\widehat\Omega_{\mathrm{III}}$ at step $n+1$ produces "memoryless"
values of partial attributes $z_{ij}^{(n+1)}$ for ACs having $L_K =
L_{\mathrm{III}}$ at step $n$ (in contrary to the
$u$\nobreakdash-\hspace{0pt}agent for ACs having $L_K =
L_{\mathrm{I}}$ at step $n$). The $z$\nobreakdash-\hspace{0pt}agent
may accelerate transitions from excited ACs to refractory ones.
This is the most important difference between our 2C model of
three-level excitable medium and the original 1C model of ZM
\cite{Zykov-1986,Loskutov-1990}.

\subsection{Refractory branch of the evolution operator}\label{subse:iii-D}

The branch $\widehat\Omega_{\mathrm{II}}$ does not produce/change
any partial attributes at all (because the intermediate level
$L_{\mathrm{II}}$ is in the state of refractority). It fulfils the
operations only over those ACs, which have $L_K = L_{\mathrm{II}}$
at step $n$:
\begin{equation}
  \varphi_{ij}^{(n+1)} =
  \widehat\Omega_{\mathrm{II}}\left(\varphi_{ij}^{(n)}\right)=
  \begin{cases}
    \varphi_{ij}^{(n)} + 1, & \text{if $\tau_e < \varphi_{ij}^{(n)} < \tau_e + \tau_r$};\\
                         0, & \text{if $\varphi_{ij}^{(n)} = \tau_e + \tau_r$}.
  \end{cases}
  \label{eq:16}
\end{equation}

Generally speaking, there are many examples of active media with
weak refractority (when the unit at the intermediate level
$L_{\mathrm{II}}$ is not absolutely isolated from its
neighborhooding units). But in this work we restrict ourselves by
the case of perfect refractority (\ref{eq:16}), which is valid,
e.~g., for the microwave phonon laser systems of the
${\mathrm{Ni^{2+}:Al_2O_3}}$ type \cite{SDM-DNM-cond-mat-0410460}.

\subsection{Reduction of the TLM model to the ZM automaton}\label{subse:iii-E}

The second-channel diffusion gives the contribution to the TLM
dynamics if $f \leq {\mathcal{N}}$, i.e. if $f \leq 4$ for $C =
C_N$, $f \leq 8$ for $C = C_M$ and so on.

At $\bigl ( (C = C_N) \wedge (f > 4) \bigr )$ our TLM model is of
ZM-like (i.e. 1C) type, and at $\bigl ( (C = C_M) \wedge (f > 8)
\bigr )$ it becomes equivalent to the original ZM model
\cite{Zykov-1986,Loskutov-1990}, which (with slight rearrangement
of cases) is as follows:
\begin{equation}
  \varphi_{ij}^{(n+1)} =
  \begin{cases}
    \varphi_{ij}^{(n)} + 1, & \text{if $0 < \varphi_{ij}^{(n)} < \tau_e + \tau_r$};\\
    1, & \text{if $\bigl( \varphi_{ij}^{(n)} = 0 \bigr) \wedge \bigl( u_{ij}^{(n+1)} \geq h \bigr)$};\\
    0, & \text{if $\left[ \bigl( \varphi_{ij}^{(n)} = 0 \bigr) \wedge \bigl( u_{ij}^{(n+1)} < h \bigr) \right] \vee \bigl( \varphi_{ij}^{(n)} = \tau_e + \tau_r \bigr)$},
  \end{cases}
  \label{eq:17}
\end{equation}
where $u_{ij}^{(n)}$ is defined as in (\ref{eq:09})--(\ref{eq:11}).

\end{widetext}

\subsection{Geometry of active media, boundary conditions and transients in TLM}\label{subse:iii-G}

Pattern evolution is very sensitive to geometry of active medium
and boundary conditions even in the framework of operation of the
same CA. There are three types of geometry and boundary condition
(GBC) combinations most commonly used in computer experiments with
CA (see Table~I). Let us consider them in details.

$\bullet$ The GBC\nobreakdash-\hspace{0pt}1 type is defined as
follows:
\begin{equation}
  S^{(n)} \bigl( i_{\mathrm{v}},j_{\mathrm{v}} \bigr) =
  S^{(n)} \bigl( i_{\mathrm{e}}^{\Gamma},j_{\mathrm{e}}^{\Gamma} \bigr),
\label{eq:18}
\end{equation}
where
\begin{equation}
  i_{\mathrm{e}}^{\Gamma} =
    \begin{cases}
    M_X             & \text{if $i_{\mathrm{v}} = 0$}; \\
    1,              & \text{if $i_{\mathrm{v}} = M_X + 1$}; \\
    i_{\mathrm{v}}, & \text{otherwise},
  \end{cases}
  \label{eq:19}
\end{equation}
\begin{equation}
  j_{\mathrm{e}}^{\Gamma} =
    \begin{cases}
    M_Y             & \text{if $j_{\mathrm{v}} = 0$}; \\
    1,              & \text{if $j_{\mathrm{v}} = M_Y + 1$}; \\
    j_{\mathrm{v}}, & \text{otherwise}.
  \end{cases}
  \label{eq:20}
\end{equation}
Here $(i_{\mathrm{v}},j_{\mathrm{v}}) \in
{\mathfrak{M}}_{\mathrm{v}}$, and the coordinates
$(i_{\mathrm{e}}^{\Gamma},j_{\mathrm{e}}^{\Gamma})$ belong to the
border set ${\mathfrak{M}}_{\mathrm{e}}^{\Gamma} \subset
{\mathfrak{M}}_{\mathrm{e}}$ of excitable area, i.~e.
$(i_{\mathrm{e}},j_{\mathrm{e}}) =
(i_{\mathrm{e}}^{\Gamma},j_{\mathrm{e}}^{\Gamma}$) if $(i = 1) \vee
(j = 1) \vee (i = M_X) \vee (j = M_Y)$.

\vspace{10pt}

$\bullet$ The GBC\nobreakdash-\hspace{0pt}2 type can be easily
defined by using an extended interval for the phase counter
$\varphi$. Let every cell $(i_{\mathrm{v}},j_{\mathrm{v}})$ in
${\mathfrak{M}}_{\mathrm{v}}$ contains one {\textit{unexcitable}}
unit. All these unexcitable units have frozen $\varphi^{(n)} \bigl(
i_{\mathrm{v}},j_{\mathrm{v}} \bigr)$ at all $n \in [0,N]$ without
reference to states of
$(i_{\mathrm{e}}^{\Gamma},j_{\mathrm{e}}^{\Gamma})$, and
\begin{equation}
  \varphi^{(n)} \bigl( i_{\mathrm{v}},j_{\mathrm{v}} \bigr) = \chi = \text{const},
  \label{eq:21}
\end{equation}
where, $\chi >  \tau_e + \tau_r$ or $\chi < 0$. These unexcitable
units in ${\mathfrak{M}}_{\mathrm{v}}$ ``absorb'' excitations at
the border of ${\mathfrak{M}}_{\mathrm{e}}$ in contrary to the case
of GBC\nobreakdash-\hspace{0pt}1, where excitations at the border
${\mathfrak{M}}_{\mathrm{e}}$ are reinjected in active medium. In
many cases this difference leads to qualitatively different
behaviour of the whole TLM. Note that by this way one may also
define TLM with {\textit{inhomogeneous}} active medium, where some
of ACs are changed to unexcitable units, i.~e. unexcitable units
are placed not only in ${\mathfrak{M}}_{\mathrm{v}}$, but in
${\mathfrak{M}}_{\mathrm{e}}$ too. This may be easily done by
introducing an additional orthogonal branch $\widehat\Omega_0$ into
the evolution operator $\widehat\Omega$. The pointed additional
branch is activated at $\varphi_{ij}^{(n)} = \chi$ only and is
simply the identity operator $\widehat\Omega_0 = \widehat{I}$,
i.e.: $\varphi_{ij}^{(n+1)} = \widehat\Omega_0 \varphi_{ij}^{(n)} =
\varphi_{ij}^{(n)}$. And even more complex behaviour of such
``impurity'' units may be defined using analogous approach (i.~e.
by introducing additional orthogonal branches in an evolution
operator): there may be pacemakers \cite{Loskutov-1990} or another
special units embedded in active medium and described by phase
counter $\varphi \neq \chi$ in extended areas $(\varphi >
\tau_e + \tau_r)$ or $(\varphi < 0)$.%

$\bullet$ The GBC\nobreakdash-\hspace{0pt}3 type is, strictly
speaking, a case of borderless system. But the starting pattern
has, of course, bounded quantity of ACs with $L_K \neq
L_{\mathrm{I}}$, located in bounded part of active medium. Hence,
during the whole evolution ($1 \leq n \leq N$) the front of growing
excited area will meet the unexcited (but excitable!) ACs only. For
the case GBC\nobreakdash-\hspace{0pt}3, there is neither absorption
of excitations at boundaries (as for
GBC\nobreakdash-\hspace{0pt}2), nor feedback by reinjection of
excitations in the active system (as for
GBC\nobreakdash-\hspace{0pt}1). From this point of view CA of
GBC\nobreakdash-\hspace{0pt}3 type are ``simpler'', than CA of
GBC\nobreakdash-\hspace{0pt}1 and GBC\nobreakdash-\hspace{0pt}2
types. On the other hand, CA of GBC\nobreakdash-\hspace{0pt}3 type
are potentially infinite discrete systems where true aperiodic
(irregular, chaotic) motions are possible, in contrast to finite
discrete systems, possessing, of course, only periodic trajectories
in certain phase space after ending of transient stage
\cite{Lorenz-etc}.

On the other hand, there is also very important difference between
GBC\nobreakdash-\hspace{0pt}2 and other two types of GBC. The
system of GBC\nobreakdash-\hspace{0pt}2 type is the single from the
three ones under consideration which interacts with the external
world dynamically (besides of relaxation). Sure enough,
GBC\nobreakdash-\hspace{0pt}1 and GBC\nobreakdash-\hspace{0pt}3 are
connected to this outer world {\textit{only}} by relaxation
channels (the $\tau_e$ and $\tau_r$ are the measures of this
connection). In contrary, a system of the
GBC\nobreakdash-\hspace{0pt}2 type interacts with surroundings
through the real boundary, which is in fact absent in toroidal
finite-size active medium of GBC\nobreakdash-\hspace{0pt}1 type and
it is absent by definition for a system of
GBC\nobreakdash-\hspace{0pt}3 type. Despite of elementary mode of
such interaction (boundary simply ''absorbs'' the outside-directed
flow of excitations from
$(i_{\mathrm{e}}^{\Gamma},j_{\mathrm{e}}^{\Gamma})$, a system of
the GBC\nobreakdash-\hspace{0pt}2 type may demonstrate very special
behaviour. The main attention in this work is devoted just to TLM
of GBC\nobreakdash-\hspace{0pt}2 type as the most realistic model
of microwave phonon laser active system, where both the mechanisms
of interaction (dynamical and relaxational) of an active medium
with the outer world are essential. Dissipation in a microwave
phonon laser active medium (highly perfect single crystal at liquid
helium temperatures) is caused by two main mechanisms: (a)
dynamical, by coherent microwave phonon and photon emission
directly through crystal boundaries and (b) relaxational, by
thermal phonon emission. There are, of course, many more or less
important differences between the TLM model and the real microwave
phonon laser systems
\cite{TPL-2001,arx-selforg-2003,TP-2004,arx04a-destab-2004}, but in
any case autonomous microwave phonon laser has only these two
mechanisms of interaction with the outer world.

\begin{table}
\begin{center}
  \caption{Geometry of active media and boundary
  conditions}
  \begin{tabular}{|c|c|c|} \hline
  {\textit{Type}} & {\textit{Geometry}} & {\textit{Boundary conditions}} \\ \hline
  GBC\nobreakdash-\hspace{0pt}1            & Bounded, Toroidal                    & Cyclic    \\ \hline
  GBC\nobreakdash-\hspace{0pt}2            & Bounded, Flat                        & Zero-flow \\ \hline
  GBC\nobreakdash-\hspace{0pt}3            & Unbounded, Flat                      & (free space) \\ \hline
  \end{tabular}
  \label{ta:01}
\end{center}
\end{table}

\subsection{Initial conditions}\label{subse:iii-H}

Patterns in ZM are described in terms of levels (or ``colors'') of
ACs, and initial conditions are formulated simply as matrix of ACs
levels $L_K$. But states of ACs in such the automata as ZM or TLM
are fully defined not only by levels $L_K$ itself. There are global
attributes which must be predefined before TLM evolution is
started. Some of partial attributes must be predefined too. These
points are essential for reproducing of the results of computer
experiments with ZM, TLM, etc.

In our work such the initial conditions for the global attributes
$\varphi_{ij}$ are used:
\begin{equation}
  \begin{cases}
    \varphi_{ij}^{(0)} = 0,          & \text{if $L_K^{(0)} = L_{\mathrm{I}}$};\\
    \varphi_{ij}^{(0)} = 1,          & \text{if $L_K^{(0)} = L_{\mathrm{III}}$};\\
    \varphi_{ij}^{(0)} = \tau_e + 1, & \text{if $L_K^{(0)} = L_{\mathrm{II}}$},
  \end{cases}
  \label{eq:22}
\end{equation}
Initial conditions for $u_{ij}$ must be defined for ground-state
AC's ($L_K^{(0)} = L_{\mathrm{I}}$) only. In this work, we suppose
\begin{equation}
  \bigl( u_{ij}^{(0)} = 0 \bigr) \quad \text{IFF} \quad
  \bigl( \varphi_{ij}^{(0)} = 0 \bigr),
  \label{eq:23}
\end{equation}
where ${\mathrm{IFF}}$ means ``if and only if''. Initial conditions
for $z_{ij}$ are undefined for all $L_K^{(0)}$ because
$z$\nobreakdash-\hspace{0pt}agent is not defined at $n=0$, see Eqn.
(\ref{eq:02}). As the result, initial conditions will be defined in
our TLM model as the matrix $ \| \varphi_{ij}^{(0)} \| $ (where $
\varphi_{ij}^{(0)} \in \{0, 1, \tau_e + 1 \} $) with additional
condition given by Eq. (\ref{eq:23}).

\section{RESULTS AND DISCUSSION}\label{se:iv}

A bounded solitary domain of excited ACs having appropriate
relaxation times and placed far from grid boundaries (or at
unbounded grid) may evolve to RSWs if and only if there is an
adjoined (but not a surrounding) domain of refractory ACs
\cite{Balakhovskii-1965,Loskutov-1990,Fisch-1991,Selfridge-1948}.
In this case RSWs apear by pair with opposite ${\mathrm{sgn}}
(Q_T)$ and integral topological charge $Q_T^{(\text{integr})}
\equiv \sum_m Q_T^{(m)}$ is obviously conserved (by infinity in
time if grid is unbounded). If such solitary excited-refractory
area is placed in a starting pattern near the grid boundary and the
GBC-2 conditions take place, the single RSW appears and evolves and
$Q_T^{(\text{integr})}$ is, of course, not conserved in this case.
These simplest and well known examples illustrate possibilities of
coexistence and competition of one or two RSWs in excitable media,
and possible scenarios of the RSW(s) evolution may be easily
forecasted.

An evolution of complex patterns with multiple, irregularly
appearing, chaotically-like drifting and colliding RSWs (or,
possibly, another spatio-temporal structures) is in essence
unpredictable without direct computing of the whole transient
stage. So, the best way to investigate cellular automaton (TLM in
particular) is to run it, because, as S.~Wolfram pointed out,
"their own evolution is effectively the most efficient procedure
for determining their future'' (see \cite{Wolfram-1985}, page 737).

Here we present results of our computer experiments with 2C model
of excitable media described in Section \ref{se:iii}. The main tool
used in these experiments was the cross-platform software package
``Three-Level Model of Excitable System''
(TLM)~\textcircled{c}~2006~S.~D.~Makovetskiy \cite{SDM-Forum-2006}.
It is based on extended and improved algotithms of discrete
modeling of three-level many-particle excitable systems, proposed
by us in \cite{SDM-Forum-2005,SDM-BScThesis-2005,SDM-Forum-2006}.
The TLM package is written in the Java 2 language
\cite{The-Java-Official-Site} using the Java 2 SDK (Standard
Edition, version 1.4.2) and the Swing Library.

A discrete system with finite set of levels may have only two types
of dynamically stable states at bounded lattice. The first of them
is stationary state, and the second is periodic one. They may be
called attractors by analogy with lumped dynamical systems (see
e.~g. \cite{Liu-2002,Wuensche-1998,Wuensche-2003}). For our
cellular automaton, the first of such the attractors is
spatially-uniform and time-independent state with
$\varphi_{ij}^{(n)} = 0$, where $(i \in [1,M_X]) \wedge (j \in
[1,M_Y])$, $n \geq n_C$. In other words, the only stationary state
of TLM is the state of full collapse of excitations at some step
$n_C$ (by definition of excitable system). The second type of TLM
attractors includes many various periodically repeated states (RSW
is a typical but not the single case). In this case
$\varphi_{ij}^{(n)} = \varphi_{ij}^{(n+T)}$ where $(i \in [1,M_X])
\wedge (j \in [1,M_Y])$, $n \geq n_P$; $n_P$ is the first step of
motion at a periodic attractor; $T$ is the integer-number period of
this motion, $T > 1$.

If starting patterns are generated with random spatial distribution
by levels, then the time intervals $n_C$ or $n_P$ may be considered
as times of full ordering in the system. Such {\textit{irregular}}
transients are of special interest from the point of view of
nonlinear dynamics of distributed systems
\cite{Awazu-2003,Crutchfield-1988,Hramov-2004,Morita-2003,Morita-2004}
because they may be bottlenecked by very slow, intermittent
morphogenesis of spatial-temporal structures. In the next
Subsections we study this collective relaxation for cases of
collapse and periodic final states of TLM evolution, including an
important case of lethargic transients.

\subsection{Collapse of RSWs}\label{subse:iv-A}

In this Subsection, we describe a typical scenario of the
two-dimensional, three-level and two-channel-diffusion (2D-3L-2C)
excitable system evolution leading to the full collapse of
excitations (when the system reaches the single point-like
spatio-temporal attractor
--- stable steady state with $L_K = L_{\mathrm{I}}$ for all ACs).
Some typical stages of such the scenario of evolution are shown at
Figure~1.

The system evolves at first stages ($0 < n \lesssim 10^4$) to a
labyrinthic structure which contains several RSW-nucleating domains
(Figure~1, $n = 10000$).

During the next stage of the evolution, the large-scale vortex is
formed (Figure~1, $n = 60000$). This vortex pushes the labirinthic
structure to the boundaries of the active medium. As the result,
the vortex occupies the whole active medium (Figure~1, $n =
100000$).

But this giant rotating structure is unstable. Its core splits into
RSWs with $Q_T = -1$, and at this stage the active medium is
filling by such the RSWs (Figure~1, $n = 160000$).

The next, relatively long stage of evolution is characterized by
restless moving and strong competition of RSWs. As the matter of
fact, this stage is of the ``winner takes all'' (WTA) type. So, at
the end of this stage, the single, fully regular RSW with $Q_T =
-1$ occupies the whole active medium (Figure~1, $n = 365000$).

But the core of the winner is not far from the active medium
boundary. Drift of this RSW leads to collision of its core with
boundary, and the core is absorbed by the latter (Figure~1, $n =
391000$). The residual nonspiral autowaves gradually run out the
active medium (Figure~1, $n = 391000$) and excitation collapses
fully at $n = n_C = 396982$ (not shown at Figure~1).

The described scenario of excitations collapse is very typical but
not the single one. Some other ways leading to collapse will be
described in a forthcoming publications. Now we will consider cases
of non-collapsing systems.

\subsection{Dynamic Stabilization, Synchronization and Coexistence of RSWs}\label{subse:iv-B}

In this Subsection, we describe a typical scenario of the 2D-3L-2C
excitable system evolution leading to dynamic stabilization and
coexistence of RSWs. In this case the system reaches one of its
periodic spatio-temporal attractors --- fully synchronized cyclic
transitions ($L_{\mathrm{I}} \rightarrow L_{\mathrm{III}}
\rightarrow L_{\mathrm{II}} \rightarrow L_{\mathrm{I}} \rightarrow
...$) for all ACs. Some typical stages of such the scenario of
evolution are shown at Figure~2.

The system evolves at first stages ($0 < n \lesssim 10^4$) to a
mixed RSW-labyrinthic structure (Figure~2, $n = 10000$), where
nucleation of RSWs has more pronounced form, than at Figure~1 (for
the same $n = 10000$.

All the subsequent stages of evolution at Figure~2 differ more and
more comparatively to Figure~1. Instead of formation of a giant
vortex, which pushes the labyrinthic structure to the boundaries of
the active medium (Figure~1), at Figure~2 we see multiple emergent
RSWs ($n = 30000$). These RSWs have different ${\mathrm{sgn}}
(Q_T)$, but RSWs with ${\mathrm{sgn}} (Q_T) = -1$ are prevailed
here (Figure~2, $n = 30000$).

As the result of such direction of evolution, at the next stage
only negative-charged, well-formed RSWs with $Q_T = - 1$ occupy the
whole active medium (Figure~2, $n = 100000$) --- instead of the
single giant multi-charged vortex with $| Q_T | > 1$ (Figure~1, $n
= 100000$).

Due to complex interactions between these RSWs as well as between
RSWs and boundaries, the processes of revival of positive-charged
RSWs take place (Figure~2, $n = 165000$). In contrast to Figure~1,
where strong competition between RSWs leads to WTA dynamics,
competition is weakened in the case under consideration (Figure~2,
$n = 204000$).

Weakening of competition and low mobility of RSWs determine
qualitatively another final (Figure~2, $n = 300000$ and $n =
987000)$ of evolution scenario (comparatively to Figure~1). Namely,
the dynamic stabilization of RSWs over the active medium is
observed at $n \approx 300000$. Configuration of the system is
precisely repeated (compare $n = 300000$ and $n = 987000$ at
Figure~2). In other words, this is spatio-temporal limit cycle.

Fully synchronized cyclic transitions for all ACs ($L_{\mathrm{I}}
\rightarrow L_{\mathrm{III}} \rightarrow L_{\mathrm{II}}
\rightarrow L_{\mathrm{I}} \rightarrow ...$) for this system
(Figure~2) is in contrast to collapse of excitations for the
previous system (Figure~1) despite of obvious transient ordering of
the spatio-temporal dynamics in the course of events for both the
cases (Figure~1 and Figure~2). Note that in {\textit{both}} these
cases we deal with spontaneous decreasing of effective freedom
degrees quantity during the evolution of the system
\cite{Haken-1983}.

Transient ordering of this kind, leading to collapse of excitations
(Figure~1), may be interpreted as self-organized degradation
(self-destruction) of the system. Final state of such the system,
of course, is not self-organized, but it is still highly ordered.
This kind of static order is known as freezing, because all ACs of
the system occupies their ground levels (like usual particles of a
physical system at absolute zero temperature).

In contrast, transient ordering, leading to dynamical stabilization
of more or less complex spatio-temporal structures (Figure~2), may
be interpreted as ``normal'' self-organization in autonomous
dissipative system \cite{Haken-1983} when most degrees of freedom
are ``slaved'' by the small rest of ones. At this self-organized
state, the system reaches the end of its evolution. But the system
remains unfrozen, it is still far from equilibrium and possesses
obvious order in the form of spatio-temporal limit cycle
(Figure~2). All parts of this system (namely RSWs) are well defined
and robust (as individual dynamical spatio-temporal
autostructures). At the same time, all these parts and entirely
synchronized at global level (as dynamical components of the whole
system demonstrating collective rhythmic motions).

Nevertheless, both these scenarios of dissipative system evolution
(self-destruction at Figure~1 and normal self-organization at
Figure~2) exhibit more or less fluent transition from initial
spatial chaos to static (Figure~1) or dynamically stable (Figure~2)
order.

A different scenario is exemplified in the next Subsection.

\subsection{Transient Chaos Caused by Competition of RSWs and Labyrinths}\label{subse:iv-C}

In this Subsection, we will describe another way to reach periodic
state in a bounded discrete 2D-3L-2C excitable system. Formally,
the final state for this new case is spatio-temporal limit cycle
too (as in the case of Figure~2). But the time of transient process
is much longer here, and the longest part of the transient proceeds
in the form of lethargic competition between various
spatio-temporal structures.

As it was already shown in the preceding Subsection (Figure~2),
spatio-temporal structures may actively interact not only between
themselves, but they interact in a nontrivial manner with
boundaries of the excitable medium too. The simplest case when
drift of RSWs leads to collisions of their cores with boundary, and
the cores are absorbed by the latter (Figure~1, $n = 391000$), is
not typical case for the system studied in this Subsection. On the
contrary, RSWs in such the system under appropriate conditions may
be "reflected" from
boundaries in a strongly nonlinear manner%
\footnote{Reflection of RSW from boundary, as well as reflection of
any other moving {\textit{dissipative}} structure (autostructure,
autowave) is usually considered as forbidden. This is correct,
commonly speaking, only in lowest (weakly nonlinear)
approximations, when the properties of surface layer are assumed as
close to properties of bulk active medium. But real processes in
surface layer may lead to strong nonlinear phenomena of, e.~g.,
revival (regeneration) of RSWs or even to their duplication,
triplication etc. Complex phenomena of such the kind may be
considered as higher-order nonlinear interactions of RSWs with
boundary, even if boundary conditions itself are ``simple'' (e.~g.
zero-flow conditions, as in the present work).}
. Moreover, due to nonlinear processes in the surface layer, the
quantity of "reflected" RSWs may exceed the quantity of primary
RSWs (phenomenon of multiplication of RSWs). A detailed description
of nonlinear "reflections" of RSWs and some accompanied phenomena
will be published in separate paper.

Here we describe a particular, but important case of highly
bottlenecked transient process, caused mainly by almost everlasting
competition of RSWs and labyrinths. At Figure~3, a fragment of
evolution is shown at $n > 10^6$ for the 2D-3L-2C system, which
differs from the previous system (Figure~2) only by two parameters:
$\tau_r = 44$; $h = 46$ for Figure~3 (instead of $\tau_r = 46$; $h
= 51$ for Figure~2). Starting pattern for the system at Figure~3 is
precisely the same as at Figure~2 (so it is not shown at Figure~3).

Previous two systems (Figures~1 and 2) reach their attractors
already at $n < 5 \cdot 10^5$. The system under consideration
(Figure~3) has very long transient time, because of lethargic
competition between RSWs and labyrinthic structures under
conditions of regeneration (nonlinear ``reflections'') and
multiplication of RSWs at boundaries. There are many fine phenomena
accompanied these competition. E.~g., the largest RSW with $Q_T =
+1$ resided in bottom left corner of the active medium (Figure~3,
$n = 1307000$) moves in the {\textit{North}} direction to the big
labyrinth. Finally, the latter absorbs the core of the pointed RSW
(Figure~3, $n = 1322000$). At the same time new small RSWs are
generated in the bulk active medium and regenerated (by nonlinear
``reflections'') in the surface layer.

The dominant drift directions of RSWs in the system under
consideration lie at
{\textit{North}}$\leftrightarrow${\textit{South}} and
{\textit{East}}$\leftrightarrow${\textit{West}} lines (this
circumstance is evident when one see a movie of the system
evolution). Drift of labyrinths is more complicated. In the
fragment shown at Figure~3, the big labyrinth is slowly moving
approximately in {\textit{North-East}} direction. A new labyrinth
is formed simultaneously. This new labyrinth pushes RSWs (Figure~3,
$n = 1322000$) etc.

The fragment of evolution shown at Figure~3 demonstrate only small
part of transient phenomena observed during this system evolution.
We will mention now such the typical stages of the system evolution
({\textit{not
shown}}%
\footnote{Additional figures to this Preprint may be requested from
the author by e-mail (see the first page of the Preprint)}
 at Figure~3):

$\bullet$ Nonlinear ``reflection'' of large RSW with $|Q_T = 1|$
from boundary accompanied by changing of ${\mathrm{sgn}} (Q_T)$ and
simultaneuos birth of several new small RSWs (satellites) nearby
the core of large RSW ($672000 \lesssim n \lesssim 687000$).

$\bullet$ Rotation of giant labyrinth around the domain occupied by
RSWs ($2060000 \lesssim n \lesssim 2150000$); erosion of this
labyrinth due to RSWs activity up to almost full dominance of RSWs
over active medium ($2160000 \lesssim n \lesssim 2180000$); revival
of labyrinth structures and resumption of competition between RSWs
and labyrinths ($2190000 \lesssim n \lesssim 2210000$).

$\bullet$ Nonlinear ``reflection'' of large RSW with $|Q_T = 1|$
from boundary with saving of ${\mathrm{sgn}} (Q_T)$ ($2309000
\lesssim n \lesssim 2324000$, upper part of active medium). Shining
examples of multiplication of RSWs at boundaries (the same interval
$2309000 \lesssim n \lesssim 2324000$, left, right and bottom
boundaries).

$\bullet$ Birth of self-organized pacemaker (a source of concentric
autowaves) due to nonlinear processes of RSW's core transformations
in the surface layer; evolution and decay of the pacemaker
($2965000 \lesssim n \lesssim 2980000$, left boundary of the active
medium).

$\bullet$ Pushing of labyrinth by a group of RSWs ($3100000
\lesssim n \lesssim 3115000$) etc.

From the formal point of view, all these stages of the system
evolution are only partial transient episodes on the long way to
the final self-organized state. We know that such the state is a
regular attractor (spatio-temporal limit cycle or at least
point-like attractor) because by definition our system is
autonomous, it has bounded quantity of discrete elements, each
element has bounded quantity of discrete states, and time is
discrete too. But we do not get concrete view of the attractor
until we reach it. Shortly, ``transient is nothing, attractor is
all''.

From an alternative point of view, each stage of the system
evolution is a day in the life of this system, and the final of
evolution is of minor interest, because no new events will come to
pass if an attractor is already reached. Shortly, from this point
of view, ``transient is all, attractor is nothing''.

But this alternative is not the single one. Another alternative is
as follows: ``transient is all {\textit{that we can see}}, because
attractor is unachievable''.

This last alternative (for a non-CA system with continuous spectrum
of states of its elementary components) was discussed in a seminal
work of J.~P.~Crutchfield and K.~Kaneko \cite{Crutchfield-1988},
which was entitled: ``Are Attractors Relevant to Turbulence?''. In
\cite{SDM-DNM-cond-mat-0410460}, this question was slightly
reformulated in the context of our studies of CA systems: ``Are
Attractors Relevant to Transient Spatio-Temporal Chaos?''
(turbulence in bounded, fully discrete system is no more than
metaphor). The answer is ``Yes'' if an attractor may be reached for
a reasonable time $\tau_{\text{attr}}$ (in fully discrete and
bounded system $\tau_{\text{attr}}$ is always limited by quantity
of all possible states of the system). But the answer is ``No'' if
$\tau_{\text{attr}}$ exceeds any possible duration of an
experiment. In this case a system with transient spatio-temporal
chaos cannot be distinguished from true chaotic system without
additional testing.

As a matter of fact, there are some intermediate classes of
phenomena ``at the edge between order and chaos'' which may appear
in bounded discrete deterministic system with large phase space.
Self-organization scenario which includes super-slow, bottlenecked,
chaotic-like stages is a signature of dominance of such an
intermediate class of system dynamics in numerical experiments.
Really, having limited time and computer capacity, one cannot reach
final self-organized state for a system with huge dimension of
phase space. Such CA (or another discrete mapping with lethargic
evolution) does not permit direct forecasting of the system future
without direct computation. So the computed part of transient
process is the single source of available information of our fully
deterministic but partially determined system.

Generally, the ideas of interconnection between chaotization,
turbulence, unpredictability (at one side) and ordering,
self-organization, long-time forecasting (at the opposite side) may
be very fruitful at least for investigation of complex
deterministic systems.

\section{CONCLUSIONS}\label{se:v}

In this work, we fulfill computer modeling of spatio-temporal
dynamics in large dicrete systems of three-level excitable ACs
interacting by short-range 2C diffusion. Computer experiments with
this 2D-3L-2C system were carried out using the cross-platform
software package ``Three-Level Model of Excitable System''
(TLM)~\textcircled{c}~2006~S.~D.~Makovetskiy \cite{SDM-Forum-2006}.

The most typical scenario of evolution is as follows. A robust RSW
or more complicated but strictly periodic in time structure is
formed being a cyclic attractor at several initial conditions
--- see Figure~2. This is an analog of limit cycle, i. e. regular
attractor known in lumped dynamical systems. Alternatively,
collapse of excitations, i.~e. full freezing of the excitable
system takes place --- see Figure~1. This is an analog of
point-like attractor in lumped dynamical systems.

Long-time evolution ($10^6 - 10^7$ iterations) of a 2D-3L-2C system
with slightly changed parameters demonstrates some unusual
phenomena including highly bottlenecked collective relaxation of
excitations. Part of these phenomena are caused by mechanism of
nonlinear regeneration of RSWs in the surface layer of an active
medium. Instead of well known scenario of RSW collapse, which takes
place after collision of RSW's core with absorbing boundary, we
observed complicated transformations of the RSWs cores leading to
nonlinear ``reflections'' of the RSWs from the normally absorbing
boundaries or even to birth of new RSWs in the surface layer.
Lethargic transient processes observed in our computer experiments
are partially caused by competitions between permanently reviving
RSWs and labyrinthic spatio-temporal structures. To our knowledge,
phenomena of such the nonlinear ``reflections'' of RSWs and
resulting die hard vorticity in excitable media with absorbing
boundaries were unknown earlier.

The author is grateful to D.~N.~Makovetskii (Institute of
Radio-Physics and Electronics, National Academy of Sciences of
Ukraine) for stimulating discussions on various aspects of this
work.

\appendix

\section{LIST OF ABBREVIATIONS}\label{ap:A}

1C, 2C, ... --- One-Channel, Two-Channel, ...

1D, 2D, ... --- One-Dimensional, Two-Dimensional, ...

2L, 3L, ... --- Two-Level, Three-Level, ...

AC --- Active Center

CA --- Cellular Automaton

CP --- Control Parameters

GBC --- Geometry and Boundary Conditions

RSW --- Rotating Spiral Wave

TLM --- Three-Level Model (of excitable system)

ZM --- Zykov-Mikhailov

\clearpage

\begin{widetext}

\begin{center}

{\textbf{FIGURE CAPTIONS}}

\vspace{10pt}

to the paper of S.~D.~Makovetskiy

\vspace{10pt}

``Numerical Modeling of Coexistence, Competition and Collapse\\ of
Rotating Spiral Waves in Three-Level Excitable Media\\ with
Discrete Active Centers and Absorbing Boundaries''

\vspace{10pt}

(figures see as separate PNG-files)

\end{center}

Figure~1: {\textbf{Birth, evolution and collapse of RSWs with the
same (by magnitude and sign) effective topological charges}}.
Dimensions of the active medium are $M_X = M_Y = 300$. Black, gray
and white pixels denote ACs in excited ($L_{\mathrm{III}}$),
refractory ($L_{\mathrm{II}}$), and ground ($L_{\mathrm{I}}$)
states respectively. Starting pattern is shown at $n = 0$. The
relaxation times of ACs are $\tau_e = \tau_r = 50$. The set of CPs
is as follows: $g = 1$; $h = 50$; $f = 3$.

\vspace{10pt}

Figure~2: {\textbf{Birth, evolution, dynamic stabilization and
coexistence of RSWs}}. Dimensions of the active medium and colors
of pixels are the same as at Figure~1. Starting pattern ($n = 0$)
{\textit{is not the same}} as at Figure~1, but statistical
properties of both patterns are almost identical. The relaxation
times of ACs and the set of control parameters are as follows:
$\tau_e = 60$; $\tau_r = 46$; $g = 1$; $h = 51$; $f = 2$.

\vspace{10pt}

Figure~3: {\textbf{Typical stage of transient spatio-temporal chaos
(competition of RSWs and labyrinthic structures) at}} $\bm{n \ggg
\max(\tau_e, \tau_r).}$ Dimensions of the active medium and colors
of pixels are the same as at Figures~1 and 2. Starting pattern
{\textit{is precisely the same}} as at Figure~2 (and not shown
here). The relaxation times of ACs and the set of CPs are as
follows: $\tau_e = 60$; $\tau_r = 44$; $g = 1$; $h = 46$; $f = 2$.

\clearpage

\end{widetext}


\begin{thebibliography}{999}

\bibitem{Awazu-2003}
 A.~Awazu and K.~Kaneko,
 Phys. Rev. Lett. {\textbf{92}}, 258302 (2004);
 Preprint
 \eprint{nlin/0310018} (2003).

\bibitem{Balakhovskii-1965}
 I.~S.~Balakhovskii,
 Biofizika (USSR) {\textbf{10}}(6), 1063 (1965),
 in Russian.

\bibitem{Bandman-2006}
 O.~L.~Bandman,
 {\textit{Cellular-Automata Models of Spatial Dynamics}},
 in: System Informatics (Siberian Branch of RAS, Novosibirsk,
 2006), Issue 10, p.57, in Russian;
 \eprint{\texttt{http://ssdonline.sscc.ru/o-l/ca-new.pdf}}

\bibitem{Bogach-1973}
 P.~G.~Bogach and L.~V.~Reshodko,
 Dopovidi Akad. Nauk Ukr. RSR (Reports of the Ukrainian Academy
 of Sciences), Series B, No.5, 442 (1973), in Ukrainian.

\bibitem{Bogach-1979}
 P.~G.~Bogach and L.~V.~Reshodko,
 {\textit{Algorithmic and Automaton Models of Smooth Muscle Operation}}
 (Naukova Dumka, Kiev, 1979), in Russian.

\bibitem{Crutchfield-1988}
 J.~P.~Crutchfield and K.~Kaneko,
 Phys. Rev. Lett. {\textbf{60}}, 2715 (1988).

\bibitem{Field-1985}
 R.~J.~Field and M.~Burger (Eds.),
 {\textit{Oscillations and Travelling Waves in Chemical Systems}}
 (Wiley, New York etc., 1985).

\bibitem{Fisch-1991}
 R.~Fisch, J.~Gravner, and D.~Griffeath,
 Statistics and Computing {\textbf{1}}, 23 (1991);\\
 \eprint{\texttt{http://psoup.math.wisc.edu/papers/tr.zip}}

\bibitem{Frank-1984}
 D.~A.~Frank-Kamenetzky,
 {\textit{Diffusion in Chemical Kinetics,
 3-rd Edition}} (Nauka, Moscow, 1987), in Russian
 (English translation of previous edition is available).

\bibitem{Giudici-Tredicce-1997}
 M.~Giudici, C.~Green, G.~Giacomelli, U.~Nespolo, and
 J.~R.~Tredicce,
 Phys. Rev. E {\textbf{55}}, 6414 (1997).

\bibitem{Gomila-2004}
 D.~Gomila, M.~A.~Mat\'ias, and P.~Colet,
 {\textit{Excitability Mediated by Localized Structures}},
 Preprint
 \eprint{nlin/0411047} (2004).

\bibitem{Hagberg-Diss-1994}
 A.~Hagberg,
 {\textit{Fronts and Patterns in Reaction-Dif\-fusion Equations}},
 Ph.~D. Thesis (Univ. of Arizona, USA, 1994);\\
 \eprint{\texttt{http://math.lanl.gov/$\sim$hagberg/Papers\\
 /dissertation/dissertation.pdf}}

\bibitem{Haken-1983}
 H.~Haken,
 {\textit{Advanced Synergetics. Instability hierarchies of
 self-organizing systems and devices}}
(Springer-Verlag, Berlin and Heidelberg, 1983).

\bibitem{Hramov-2004}
 A.~E.~Hramov, A.~E.~Khramova, I.~A.~Khromova, and A.~A.~Koronovskii,
 Nonlin. Phenom. Complex Syst. {\textbf{7}}(1), 1 (2004).

\bibitem{The-Java-Official-Site}
 {\textit{The Java Official Site}},
 \eprint{\texttt{http://java.sun.com/}}

\bibitem{Kapral-1995}
 R.~Kapral, R.~Showalter (Eds.),
 {\textit{Chemical Waves and Patterns}},
 (Kluwer Academic Press, Dortrecht, 1995).

\bibitem{Kolmogorov-etc}
 A.~N.~Kolmogorov,
 Uspekhi Mathem. Nauk (USSR) {\textbf{8}}(4), 175 (1953), in Russian;
 %
 A.~N.~Kolmogorov and V.~A.~Uspenskii,
 Uspekhi Mathem. Nauk (USSR) {\textbf{13}}(4), 3 (1958), in Russian;
 %
 V.~A.~Uspenskii and A.~L.~Semenov,
 {\textit{Theory of Algorithms}}
 (Nauka, Moscow, 1987),
 in Russian.

\bibitem{Liu-2002}
 R.~F.~Liu and C. C. Chen,
 Preprint
 \eprint{nlin/0209005} (2002).

\bibitem{Lo-2003}
 C.-P.~Lo, N.~S.~Nedialkov, and J.-M.~Yuan,
 Preprint
 \eprint{math/0307394} (2003).

\bibitem{Lorenz-etc}
 E.~N.~Lorenz,
 Physica D {\textbf{35}}, 299 (1989);
 %
 J.~L.~McCauley,
 Phys. Scripta {\textbf{20}}, 1 (1990);
 %
 J.~Palmore and C.~Herring,
 Physica D {\textbf{42}}, 99 (1990);
 %
 P.~M.~B.~Vitanyi,
 Preprint
 \eprint{cond-mat/0303016} (2003);
 %
 Masato Ida and Nobuyuki Taniguchi,
 Phys. Rev. E {\textbf{68}}, 036705 (2003);
 %
 Masato Ida and Nobuyuki Taniguchi,
 Phys. Rev. E {\textbf{69}}, 046701 (2004).

\bibitem{Loskutov-1990}
 A.~Yu.~Loskutov and A.~S.~Mikhailov,
 {\textit{Introduction to Synergetics}} (Nauka, Moscow, 1990),
 in Russian.

\bibitem{DNM-Diss-1983}
 D.~N.~Makovetskii,
 {\textit{Dissertation}}
 (Inst. of Radio-Phys. and Electron., Ukrainian Acad. Sci.,
 Kharkov, 1983);
 %
 {\textit{Diss. Summary}} (Inst. of Low
 Temper. Phys. and Engin., Ukrainian Acad. Sci., Kharkov, 1984),
 in Russian.

\bibitem{TPL-2001}
 D.~N.~Makovetskii,
 Tech. Phys. Lett. {\textbf{27}}(6), 511 (2001)
 [translated from: Pis'ma Zh. Tekh. Fiz. {\textbf{27}}(12), 57
 (2001)];\\
 \eprint{\texttt{http://www.ioffe.rssi.ru/journals/pjtf/2001/12\\/p57-64.pdf}}

\bibitem{arx-selforg-2003}
 D.~N.~Makovetskii,
 {\textit{Self-Organization in Multimode Microwave Phonon Laser (Phaser):
 Experimental Observation of Spin-Phonon Cooperative Motions}},
 Preprint
 \eprint{cond-mat/0303188} (2003).

\bibitem{TP-2004}
 D.~N.~Makovetskii,
 Tech. Phys. {\textbf{49}}(2), 224 (2004);\\
 see also \eprint{\texttt{http://dx.doi.org/10.1134/1.1648960}}
 [translated from: Zh. Tekh. Fiz. {\textbf{74}}(2), 83 (2004);\\
 \eprint{\texttt{http://www.ioffe.rssi.ru/journals/jtf/2004/02\\/p83-91.pdf}}~]

\bibitem{arx04a-destab-2004}
 D.~N.~Makovetskii,
 {\textit{Superslow Self-Organized Motions in a Multimode Microwave
 Phonon Laser (Phaser) under Resonant Destabilization of Stationary
 Acoustic Stimulated Emission}},
 Preprint
 \eprint{cond-mat/0402640} (2004).

\bibitem{UFZh-2006}
 D.~N.~Makovetskii,
 {\textit{Slowing-Down of Transient Processes of Fine Structure
 Formation in Power Spectra of Microwave Phonon Laser (Phaser)}},
 Ukrainian Journ. Phys. (2006), to be published.

\bibitem{SDM-DNM-cond-mat-0410460}
 S.~D.~Makovetskiy and D.~N.~Makovetskii,
 {\textit{A Computational Study of Rotating Spiral Waves
 and Spatio-Temporal Transient Chaos in a Deterministic
 Three-Level Active System}},
 Preprint
 \eprint{cond-mat/0410460}, version~2 (2005).

\bibitem{SDM-Forum-2005}
 S.~D.~Makovetskiy,
 {\textit{Program for Modeling of Spatio-Temporal Structures in
 Three-Level Lasers,}} in: Proc. of the 9-th Intl. Forum
 ``Radioelectronics and Youth in the XXI Century''
 (KNURE, Kharkiv, 2005), p.348, in Russian.

\bibitem{SDM-BScThesis-2005}
 S.~D.~Makovetskiy,
 {\textit{Software for Modeling of Spatio-Temporal Structures
 in Three-Level Lasers}}, B.~Sc. Thesis (Kharkiv National
 University of Radio Electronics, Ukraine, 2005), in Russian.

\bibitem{SDM-Forum-2006}
 S.~D.~Makovetskiy,
 {\textit{A Method of Numerical Modeling of Non-Stationary Processes in
 Three-Level Excitable Media and its Software Implementation by the
 Java Language,}} in: Proc. of the 10-th Intl. Forum
 ``Radioelectronics and Youth in the XXI Century''
 (KNURE, Kharkiv, 2006), in Russian.

\bibitem{Medvinskii-2002}
 A.~B.~Medvinskii, I.~A.~Tikhonova, D.~A.~Tikhonov, G.~R.~Ivanitskii,
 S.~V.~Petrovskii, B.-L.~Li, E.~Venturino, and H.~Malshow,
 Physics - Uspekhi {\textbf{45}}(1), 27 (2004)
 [translated from: Uspekhi Fiz. Nauk {\textbf{172}}(1),
 31 (2002)].

\bibitem{Morita-2003}
 H.~Morita and K.~Kaneko,
 Europhys. Lett. {\textbf{66}}, 198 (2004);
 Preprint
 \eprint{cond-mat/0304649} (2003).

\bibitem{Morita-2004}
 H.~Morita and K.~Kaneko,
 Preprint
 \eprint{nlin/0407460}\\ (2004).

\bibitem{Murray-2002}
 J.~D.~Murray,
 {\textit{Mathematical Biology, Volumes 1--2,}} 3-rd Edition (Springer,
 2002).

\bibitem{Neumann-1966}
 J.~von~Neumann,
 {\textit{Theory of Self-Reproducing Automata}}, edited and
 completed by A.~W.~Burks
 (Illinois Univ. Press, Urbana \& London, 1966).

\bibitem{Oya-2005}
 T.~Oya, T.~Asai, T.~Fukui, and Y,~Amemiya,
 Int. J. Unconvential Computing {\textbf{1}}, 177 (2005).

\bibitem{Packard-2004}
 N.~Packard and R.~Shaw,
 Preprint
 \eprint{cond-mat/0412626} (2004).

\bibitem{Piwonski-etc}
 %
 T.~Piwonski, J.~Houlihan, Th.~Busch, and G.~Huyet,
 {\textit{Delay Induced Excitability}},
 Preprint
 \eprint{cond-mat/0411385} (2004);
 %
 J.~Houlihan, D.~Goulding, Th.~Busch, C.~Masoller, and G.~Huyet,
 Phys. Rev. Lett. {\textbf{92}}, 050601 (2004);
 %
 D.~Curtin, S.~P.~Hegarty, D.~Goulding, J.~Houlihan, Th.~Busch,
 C.~Masoller, and G.~Huyet,
 Phys. Rev. E {\textbf{70}}, 031103 (2004).

\bibitem{Provata-2003}
 A.~Provata and G.~A.~Tsekouras,
 Phys. Rev. E {\textbf{67}}, 056602 (2003).

\bibitem{Reshodko-etc}
 L.~V.~Reshodko,
 Kybernetika (Prague) {\textbf{10}}(5), 409 (1974);
 %
 L.~V.~Reshodko and S.~Bures,
 Biol. Cybern. {\textbf{18}}(3/4), 181 (1975);
 %
 L.~V.~Reshodko and Z.~Drska,
 J. Theor. Biol. {\textbf{69}}(4), 568 (1977).

\bibitem{Selfridge-1948}
 O.~Selfridge,
 Arch. Inst. Cardiologia de Mexico {\textbf{53}}, 113 (1948).

\bibitem{Staliunas-1995}
 K.~Staliunas and C.~O.~Weiss,
 J. Opt. Soc. Amer. B {\textbf{12}}, 1142 (1995).

\bibitem{Szabo-2002}
 G.~Szab\'o and A.~Szolnoki,
 Phys. Rev. E {\textbf{65}}, 036115 (2002).

\bibitem{Tredicce-1985}
 J.~R.~Tredicce, F.~T.~Arecchi, G.~L.~Lippi, and G.~P.~Puccioni,
 J. Opt. Soc. Amer. B {\textbf{2}}, 173 (1985).

\bibitem{Tyson-1985}
 J.~J.~Tyson,
 in {\textit{Oscillations and Travelling Waves in Chemical
 Systems}}, ed. by R.~J.~Field and M.~Burger (Wiley, New York etc., 1985).

\bibitem{Vanag-2004}
 V.~K.~Vanag,
 Physics - Uspekhi {\textbf{47}}(12), 1177 (2004)
 [translated from: Uspekhi Fiz. Nauk (Russia) {\textbf{174}}(9),
 991 (2004)].

\bibitem{Weiss-1993}
 C.~O.~Weiss, H.~R.~Telle, K.~Staliunas, and M.~Brambilla,
 Phys. Rev. A. {\textbf{47}}, R1616 (1993).

\bibitem{Weiss-1999}
 C.~O.~Weiss, M.~Vaupel, K.~Staliunas, G.~Slekys, and V.~B.~Taranenko,
 Appl. Phys. B {\textbf{68}}, 151 (1999).

\bibitem{Weiss-2004}
 C.~O.~Weiss,
 in: {\textit{Proc. 6-th Intl. Conf. LFNM'2004}} (V.~N.~Karazin National
 University \& National University of Radio Electronics, Kharkov,
 2004), pp.~207-208.

\bibitem{Wiener-1946}
 N.~Wiener and A.~Rosenblueth,
 Arch. Inst. Cardiologia de Mexico  {\textbf{16}}(3-4), 205 (1946).

\bibitem{Wolfram-1985}
 S.~Wolfram,
 Phys. Rev. Lett. {\textbf{54}}, 735 (1985).

\bibitem{Wolfram-2002}
 S.~Wolfram,
 {\textit{A New Kind of Science}}
 (Wolfram Media, Champaign, 2002);
 \eprint{\texttt{http://www.wolframscience.com/}}

\bibitem{Wuensche-1998}
 A.~Wuensche,
 in {\textit{Complex Systems '98}}
 (Univ. of New South Wales, Sydney, 1998);\\
 \eprint{\texttt{ftp://ftp.cogs.susx.ac.uk/pub/users/andywu\\/papers%
 /complex98.ps.gz}}

\bibitem{Wuensche-2003}
 A.~Wuensche,
 Discrete Dynamics Lab (DDLab): Software Package, ver. m04
 (Discrete Dynamics, Santa Fe, 2003);
 \eprint{\texttt{http://www.ddlab.com}}

\bibitem{Zykov-1986}
 V.~S.~Zykov and A.~S.~Mikhailov,
 Sov. Phys. -- Doklady {\textbf{31}}(1), 51 (1986)
 [translated from: Dokl. Acad. Nauk SSSR {\textbf{286}}(2), 341 (1986)].

\end{thebibliography}
\end{document}